\documentclass[aps,prb,twocolumn,groupedaddress,nopacs]{revtex4}
\usepackage{graphicx}
\usepackage{dcolumn}
\usepackage{bm}
\usepackage{amssymb}
\usepackage{color}

\begin{document}


\title{Formation of polyhedral vesicles and polygonal membrane tubes induced by banana-shaped proteins} 

\author{Hiroshi Noguchi}
\email[]{noguchi@issp.u-tokyo.ac.jp}
\affiliation{
Institute for Solid State Physics, University of Tokyo,
 Kashiwa, Chiba 277-8581, Japan}

\date{\today}

\begin{abstract}
The shape transformations of fluid membranes induced by curved protein rods are studied using meshless membrane simulations. 
The rod assembly at low rod density induces a flat membrane tube and oblate vesicle.
It is found that the polyhedral shapes are stabilized at high rod densities.
The discrete shape transition between triangular and buckled discoidal tubes is obtained
and their curvature energies are analyzed by a simple geometric model.
For vesicles, triangular hosohedron and elliptic-disk shapes are formed in equilibrium, 
whereas tetrahedral and triangular prism shapes are obtained as metastable states.
\end{abstract}


\maketitle

\section{Introduction}

Single-component fluid vesicles
exhibit various morphologies such as stomatocyte, pear, and starfish shapes.
These shapes can be understood in terms of the membrane bending free energy with area-difference elasticity.\cite{lipo95,seif97,canh70,helf73,svet89,saka14,nogu15a} 
A red-blood-cell shape, discocyte, can be obtained by the energy minimization of the bending energy with fixed area and volume.
However, many cell organella have much more complicated shapes depending on their functions,
and their shapes cannot be reproduced by single-component lipid membranes.
For example, the Golgi apparatus consists of stacks of discoidal membranes.
These membrane shapes are regulated by various proteins.\cite{zimm06,baum10,joha14,ross14}
However, the mechanism of the shape regulation has not been clarified.

In particular, 
protein domains called
 BAR (Bin/Amphiphysin/Rvs) domains~\cite{itoh06,masu10,mim12a,suar14}
have recently been receiving increased attention experimentally and theoretically.
The BAR domains consist of a banana-shaped dimer and
 are categorized into subsets of unique families 
including ``classical'' BARs, F-BARs (Fes/CIP4 homology-BAR), and I-BARs (Inverse-BAR).
The BAR domain absorbs onto the biomembrane and changes the local curvature.
The extension of membrane tubes from liposomes and the formation of a cylindrical scaffold
have been experimentally observed.\cite{itoh06,masu10,mim12a,fros08,sorr12,zhu12,tana13,shi15}
Dysfunctions of the BAR proteins are considered to be implicated in neurodegenerative, cardiovascular, and neoplastic diseases.

Membrane tubulation and budding can be generated by the adhesion of objects inducing isotropic spontaneous curvature 
such as spherical colloids and conical transmembrane molecules.\cite{lipo13} 
In the last decade, the membrane-mediated interactions between the isotropic adhesives were intensively investigated.\cite{reyn07,atil07,auth09,sari12}
In contrast, the BAR domains are banana-shaped and generate an anisotropic curvature.
This anisotropic spontaneous curvature is considered to give pronounced effects to control local membrane shapes.
Recently, the classical Canham--Helfrich curvature free energy \cite{canh70,helf73} 
was extended to anisotropic curvatures.\cite{four96,kaba11,igli06}
The adsorption and assembly of the BAR domains were investigated using atomic and coarse-grained molecular simulations.\cite{arkh08,yu13,simu13,simu15}
Formation of discoidal vesicles and tubulation were simulated using a dynamically triangulated membrane model \cite{rama12,rama13} 
and meshless membrane models.\cite{ayto09,nogu14,nogu15}
Despite these numerous advancements, 
the physics of membrane shape deformation caused by anisotropic curvature is not completely understood.

In this paper, we describe shape transformations of membrane tubes and vesicles using an implicit-solvent meshless membrane model.\cite{nogu09,nogu06,shib11,nogu14,nogu15}
In particular, we focus on the membrane shapes at high protein densities.
A BAR domain is modeled as a banana-shaped rod,
and it is assumed to be strongly adsorbed onto the membrane.
In order to investigate the membrane-curvature-mediated interactions,
no direct attractive interaction is considered between the rods.
The membrane-mediated interactions induce the assembly of the protein rods.\cite{nogu14}
We will show the assembly of protein rods induces the formation of polygonal tubes and polyhedral vesicles.
Polyhedral vesicles were previously reported 
for gel-phase membranes~\cite{sack94},  phase-separated membranes~\cite{veat03,gudh07,hu11,nogu12a},
and fluid membranes with the accumulation of specific lipids or defects on the polyhedral edges or vertices.\cite{dubo01,nogu03,hase10}
Here, we demonstrate that the polyhedral vesicles can be stabilized by the protein rods.

In Sec.~\ref{sec:method}, the simulation model and method are provided.
In  Sec.~\ref{sec:tube},  the shape transformations of the membrane tubes induced by the protein rods are described.
In  Sec.~\ref{sec:geo}, a simple geometric model is proposed to understand the results of Sec.~\ref{sec:tube}.
In  Sec.~\ref{sec:ves}, the shape transformations of vesicles are described.  
The summary and discussion are given in Sec.~\ref{sec:sum}.

\section{Simulation Model and Method}\label{sec:method}

The details of the meshless membrane model and protein rods are described in Ref.~\onlinecite{shib11} and Ref.~\onlinecite{nogu14}, respectively.
Here, we briefly describe the model.
A fluid membrane is represented by a self-assembled one-layer sheet of $N$ particles.
The position and orientational vectors of the $i$-th particle are ${\bf r}_{i}$ and ${\bf u}_i$, respectively.
The membrane particles interact with each other via a potential $U=U_{\rm {rep}}+U_{\rm {att}}+U_{\rm {bend}}+U_{\rm {tilt}}$.
The potential $U_{\rm {rep}}$ is an excluded volume interaction with a diameter $\sigma$ for all pairs of particles.
Solvent is implicitly accounted for by an effective attractive potential 
 $U_{\rm {att}}$. It is a pairwise attractive potential at low particle density while the attraction is
smoothly truncated at high density.
This truncation allows formation of a fluid membrane in a wide range of parameters.
The bending and tilt potentials
are given by 
 $U_{\rm {bend}}=(k_{\rm {bend}}/2) \sum_{i<j} ({\bf u}_{i} - {\bf u}_{j} - C_{\rm {bd}} \hat{\bf r}_{i,j} )^2 w_{\rm {cv}}(r_{i,j})$
and $U_{\rm {tilt}}=(k_{\rm{tilt}}/2) \sum_{i<j} [ ( {\bf u}_{i}\cdot \hat{\bf r}_{i,j})^2
 + ({\bf u}_{j}\cdot \hat{\bf r}_{i,j})^2  ] w_{\rm {cv}}(r_{i,j})$, respectively,
where ${\bf r}_{i,j}={\bf r}_{i}-{\bf r}_j$, $r_{i,j}=|{\bf r}_{i,j}|$,
 $\hat{\bf r}_{i,j}={\bf r}_{i,j}/r_{i,j}$, and $w_{\rm {cv}}(r_{i,j})$ is a weight function.
The spontaneous curvature $C_0$ of the membrane is 
given by $C_0\sigma= C_{\rm {bd}}/2$. \cite{shib11}
In this study, $C_0=0$ except for the membrane particles belonging to the protein rods.

The protein rod is modeled as a linear chain of $N_{\rm {sg}}$ membrane particles.
We use $N_{\rm {sg}}=10$, which corresponds to the typical aspect ratio of the BAR domains.
The BAR domain width is approximately $2$ nm, and the length ranges from $13$ to $27$ nm. \cite{masu10}
The protein rod has spontaneous curvature $C_{\rm {rod}}$ along the rod axis
and zero spontaneous curvature perpendicular to the rod axis.
Hereafter, we call the membrane particle consisting of the protein rod a protein particle.
The protein particles in each protein rod
 are connected by a bond potential $U_{\rm {rbond}}/k_{\rm B}T = (k_{\rm {rbond}}/2\sigma^2)(r_{i+1,i}-l_{\rm rod})^2$.
The bending potential is given by $U_{\rm {rbend}}/k_{\rm B}T = (k_{\rm {rbend}}/2)(\hat{\bf r}_{i+1,i}\cdot\hat{\bf r}_{i,i-1}- C_{\rm r})^2$,
where $C_{\rm r}=1- (C_{\rm {rod}}l_{\rm rod})^2/2$.
For bonded pairs of protein particles, the corresponding spontaneous curvature and
 four times larger values of the coefficients of $U_{\rm {bend}}$ and $U_{\rm {tilt}}$ are
employed to prevent the rod from bending tangentially to the membrane.

We employ the parameter sets used in Ref.~\onlinecite{nogu14}.
The membrane has mechanical properties that are typical for lipid membranes:
the bending rigidity $\kappa/k_{\rm B}T=15 \pm 1$,
the area of the tensionless membrane per particle $a_0/\sigma^2=1.2778\pm 0.0002$,
the area compression modulus $K_A\sigma^2/k_{\rm B}T=83.1 \pm 0.4$,
and
the edge line tension $\Gamma\sigma/k_{\rm B}T= 5.73 \pm 0.04$,
where $k_{\rm B}T$ denotes the thermal energy.
This edge tension $\Gamma$ is sufficiently large to prevent membrane rupture in this study.
Molecular dynamics with a Langevin thermostat is employed.\cite{shib11,nogu11}
In addition to canonical ensemble simulations,
replica exchange molecular dynamics \cite{huku96,okam04} for the rod curvature $C_{\rm {rod}}$ \cite{nogu14} 
is used to obtain the thermal equilibrium states.
The simulation data of the replica exchange method (canonical simulations) are indicated by 
 lines (the symbols with lines) in the figures.
The error bars are estimated from three or four independent runs.
The results are displayed with the rod length $l_{\rm {rod}}=10\sigma$ for the length unit and
 $k_{\rm B}T$ for the energy unit.

The rod assembly on
membrane tubes with $N=2400$ and vesicles with $N=2400$ and $9600$ were investigated
at the rod density $\phi_{\rm {rod}}=N_{\rm {rod}}N_{\rm {sg}}/N=0.1$ to $0.5$, 
where $N_{\rm {rod}}$ is the number of the rods.
The tube length is fixed as $L_z=4.8r_{\rm {rod}}$ in the longitudinal ($z$) direction.
The periodic boundary conditions are employed.
The radius of the tube is $R_{\rm {cyl}}= 0.989 r_{\rm {rod}}$;
the radii of the vesicles
are $R_{\rm {ves}}= 1.54 r_{\rm {rod}}$ and $3.07 r_{\rm {rod}}$ at $N=2400$ and $9600$, respectively, in the absence of the rods.

\section{Membrane tubes}
\label{sec:tube}

A membrane tube extended from a vesicle by optical tweezers and micropipette 
is a typical experimental setup to investigate protein sorting onto lipid membranes.\cite{sorr12,zhu12,shi15}
In our previous paper~\cite{nogu14},
we reported the coupling between self-assembly of the protein rods and the shape deformation of the membrane tube
at low rod densities from $\phi_{\rm {rod}}=0.1$ to $0.25$.
First, we review the rod assembly at low rod density and then
 describe the behavior at higher density.

\begin{figure}
\includegraphics{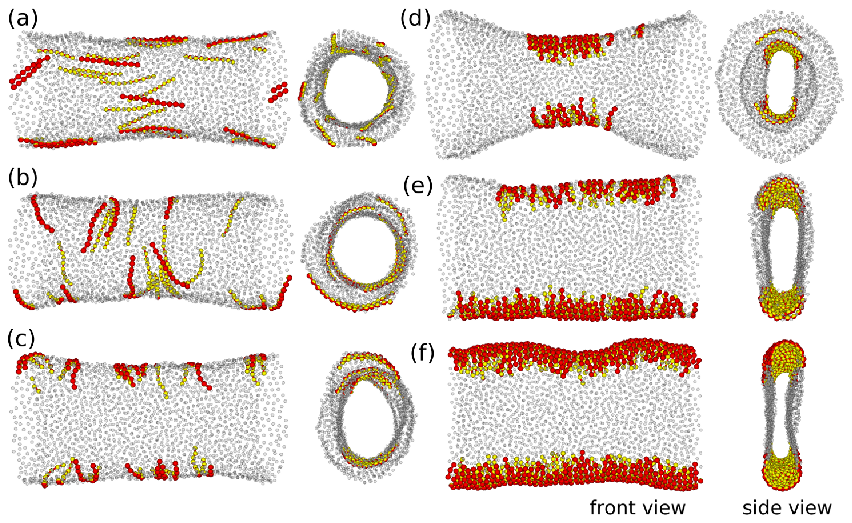}
\caption{
Snapshots of a membrane tube with protein rods.
(a) $C_{\rm {rod}}=0$ and $\phi_{\rm {rod}}=0.1$.
(b) $C_{\rm {rod}}r_{\rm {rod}}=2$  and $\phi_{\rm {rod}}=0.1$.
(c) $C_{\rm {rod}}r_{\rm {rod}}=3.2$  and $\phi_{\rm {rod}}=0.1$.
(d) $C_{\rm {rod}}r_{\rm {rod}}=3.75$  and $\phi_{\rm {rod}}=0.1$.
(e) $C_{\rm {rod}}r_{\rm {rod}}=3.75$  and $\phi_{\rm {rod}}=0.25$.
(f) $C_{\rm {rod}}r_{\rm {rod}}=3.75$  and $\phi_{\rm {rod}}=0.4$.
A protein rod is displayed as
a chain of spheres whose halves are colored
in red and in yellow.
The orientation vector ${\bf u}_i$ lies along the direction from the 
yellow to red hemispheres.
Transparent gray particles represent
membrane particles.
}
\label{fig:snap_cylrho}
\end{figure}

\begin{figure}
\includegraphics{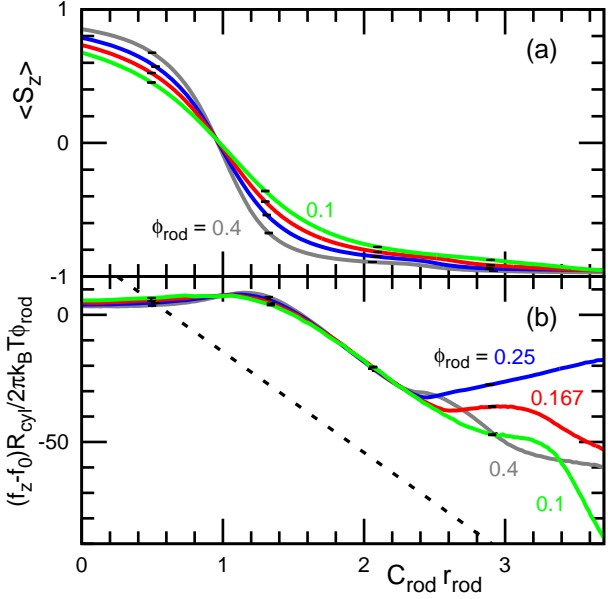}
\caption{ 
Rod curvature $C_{\rm {rod}}$ dependence of  
(a) the mean orientation degree $S_z$ and (b) the mean axial force $f_z$ of the membrane tube 
for $\phi_{\rm {rod}}=0.1$, $0.167$, $0.25$, and $0.4$.
The dashed line is obtained from Eq.~(\ref{eq:fz2}) with $\kappa_1=40k_{\rm B}T$.
Error bars are displayed at several data points.
}
\label{fig:vir_rho}
\end{figure}

\begin{figure}
\includegraphics{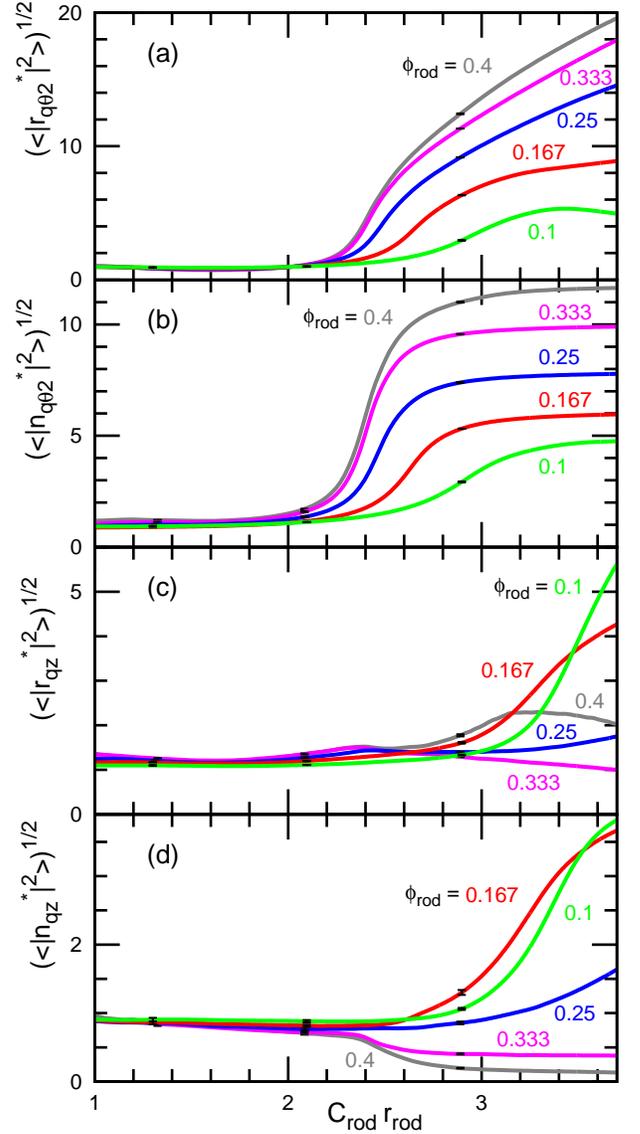}
\caption{ 
Rod curvature $C_{\rm {rod}}$ dependence of  
the mean amplitudes of (a), (c) shape deformation and (b), (d) rod densities of membrane tubes 
for $\phi_{\rm {rod}}=0.1$, $0.167$, $0.25$, and $0.4$.
The amplitudes of the lowest Fourier mode along the azimuthal ($\theta$) and longitudinal ($z$) directions
are calculated for the membrane shapes ($r_{q\theta 2}$ and $r_{qz}$) and densities ($n_{q\theta 2}$ and $n_{qz}$) 
of the center of mass of the protein rods.
The Fourier amplitudes are normalized by those at $C_{\rm {rod}}=0$ (denoted by $*$).
Error bars are displayed at several data points.
}
\label{fig:r_nrho}
\end{figure}

At small rod curvatures for $0 \leq C_{\rm {rod}} r_{\rm {rod}}\lesssim 2$, 
the rods are randomly distributed on the membrane [see Figs. \ref{fig:snap_cylrho}(a),(b)].
As the rod curvature $C_{\rm {rod}}$ increases,
the rod orientation changes from the tube ($z$) axis direction into the azimuthal ($\theta$) direction,
so that the orientational order parameter $S_z = (1/N_{\rm rod})\sum_i ( 2{s_{i,z}}^2-1 )$
decreases,
where $s_{i,z}$ is the $z$ component of the orientation vector of the $i$-th rod
[see Fig. \ref{fig:vir_rho}(a)].
With increasing $\phi_{\rm {rod}}$,
this orientation change is slightly enhanced by rod--rod interactions.

In a cylindrical tube of a homogeneous membrane, an axial force
\begin{equation}
f_z = \frac{\partial F}{\partial L_z}\bigg|_A = 2\pi\kappa \Big(\frac{1}{R_{\rm cyl}} - C_0\Big), \label{eq:fz1}
\end{equation}
is generated by the membrane bending energy,
since an increase in the axial length results in a decrease in the cylindrical radius.\cite{shib11}
Figure \ref{fig:vir_rho}(b) shows the force $f_z-f_0$ induced by the rods with normalization by $\phi_{\rm {rod}}$,
where $f_0= 102 k_BT/r_{\rm {rod}}$ is the force in the absence of the rods.
All curves in Fig. \ref{fig:vir_rho}(b) overlap for $0 \leq C_{\rm {rod}} r_{\rm {rod}}\lesssim 2$.
Thus, the effects of the rods on the axial force are proportional to the rod density $\phi_{\rm {rod}}$
unless the rods begin to self-assemble.
During the rod orientation change for $0 \leq C_{\rm {rod}} r_{\rm {rod}}\lesssim 1.2$,
the force $f_z$ is almost constant, so that it does not contribute to the effective spontaneous curvature of the membrane.
After the orientation is almost saturated ($1.2 \lesssim C_{\rm {rod}} r_{\rm {rod}}\lesssim 2$), 
$f_z$ linearly decreases with increasing $C_{\rm {rod}}$.
We will discuss the slope of this decrease later in Sec.~\ref{sec:geo}.

\begin{figure}
\includegraphics[width=6.5cm]{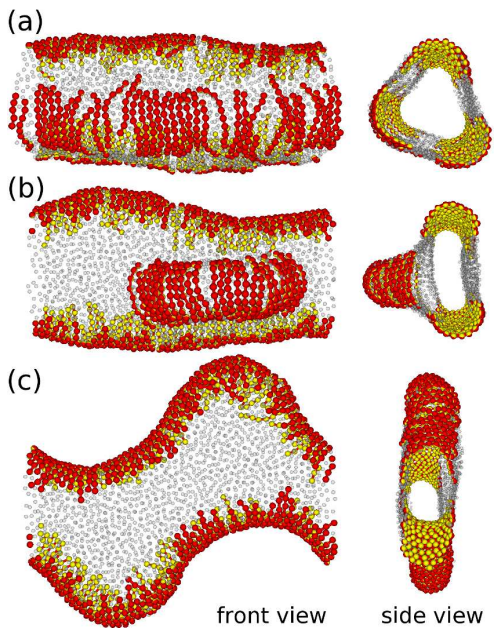}
\caption{ 
Snapshots of membrane tubes at $\phi_{\rm {rod}}=0.5$.
(a) Triangular tube at $C_{\rm {rod}}r_{\rm {rod}}=3$.
(b) Triangular tube at $C_{\rm {rod}}r_{\rm {rod}}=3.4$.
One of the edges of the triangle is partially broken and
 a discoidal bud is formed.
(c) Buckled discoidal tube at $C_{\rm {rod}}r_{\rm {rod}}=3.25$.
}
\label{fig:snap_n120}
\end{figure}

\begin{figure}
\includegraphics{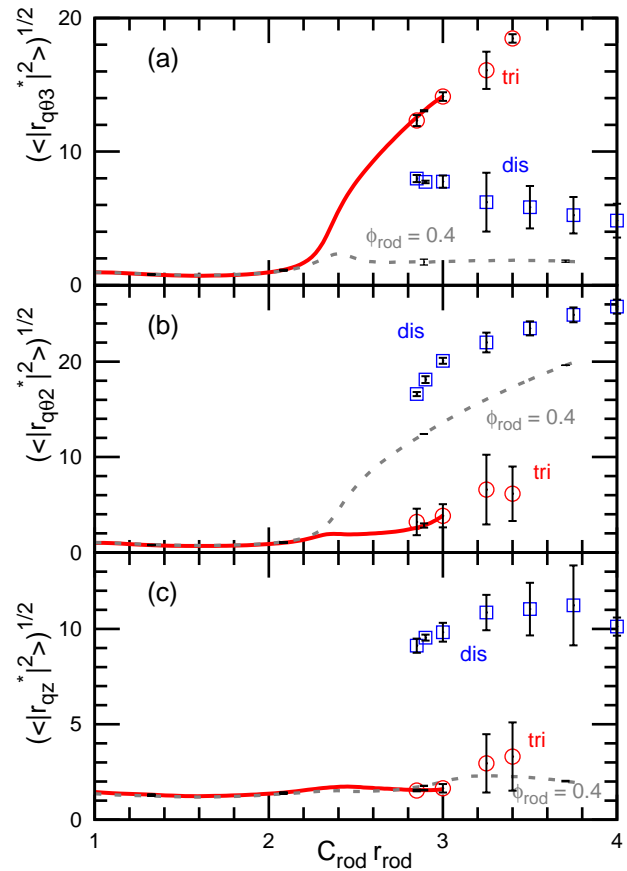}
\caption{ 
Fourier amplitudes of shape deformation of membrane tubes.
The solid lines and symbols represent the results of the replica exchange simulation and canonical simulations, 
respectively, at $\phi_{\rm {rod}}=0.5$.
The discoidal and triangular tubes are indicated by 'dis' and 'tri', respectively.
The dashed lines represent the results of the replica exchange simulation at $\phi_{\rm {rod}}=0.4$.
Error bars are displayed at several data points for the replica exchange simulations.
}
\label{fig:r_n120}
\end{figure}

As $C_{\rm {rod}}$ increases further ($2 \lesssim C_{\rm {rod}} r_{\rm {rod}}\lesssim 3$),
the rods assemble in the azimuthal ($\theta$) direction and the tube deforms into an elliptical shape 
[see Fig. \ref{fig:snap_cylrho}(c)].
As $C_{\rm {rod}}$ increases even further,
the rods also assemble in the longitudinal ($z$) direction [see Fig. \ref{fig:snap_cylrho}(d)].
This two-step assembly is captured by the amplitudes of the Fourier modes (see Fig. \ref{fig:r_nrho}).
The lowest Fourier modes of the membrane shape and rod density  along the longitudinal ($z$) direction
are given by  $r_{qz}= (1/N)\sum_i r_i \exp(-2\pi z_i {\rm i}/L_z)$
and  $n_{qz}= (1/N_{\rm {rod}})\sum_i \exp(-2\pi z_i {\rm i}/L_z)$, respectively,
where  $r_i^2 = x_i^2+ y_i^2$.
In the azimuthal ($\theta$) direction,
the $m$-th Fourier modes are 
$r_{q\theta m}= (1/N)\sum_i r_i \exp(-m \theta_i {\rm i})$
and
$n_{q\theta m}= (1/N_{\rm {rod}})\sum_i \exp(-m \theta_i {\rm i})$,
where $\theta_i=\tan^{-1}(x_i/y_i)$.
The modes of $m=2$ and $m=3$ indicate elliptical and triangular deformations, respectively.
The amplitudes of the membrane shape $r_{q\theta 2}$ and rod density $n_{q\theta 2}$ 
along the $\theta$ direction increase together,  
and subsequently, the amplitudes of $r_{qz}$ and $n_{qz}$ along the $z$ direction increase at $\phi_{\rm {rod}}=0.1$.

As $\phi_{\rm {rod}}$ increases,
azimuthal assembly occurs at smaller values of $C_{\rm {rod}}$ [see Figs. \ref{fig:r_nrho}(a), (b)].
In contrast, longitudinal assembly is suppressed at $\phi_{\rm {rod}} \gtrsim 0.3$.
This is caused by  too high density of the rods at the disk edges of the tubes [see Figs.~\ref{fig:snap_cylrho}(e), (f)].
At $\phi_{\rm {rod}}=0.4$, the edges are completely filled by the rods.
The small values of $<|n_{qz}^*|^2>$ indicate that little space is left for the rods to move translationally along the edges.

As $\phi_{\rm {rod}}$ increases further,
the phase behavior drastically changes.
At $\phi_{\rm {rod}}=0.5$, a triangular tube is formed with increasing $C_{\rm {rod}}$ 
(see Figs.~\ref{fig:snap_n120} and \ref{fig:r_n120}).
With further increasing $C_{\rm {rod}}$,
the triangular shapes become unstable so that a buckled discoidal tube is formed [see Fig.~\ref{fig:snap_n120}(c)].
At the largest values of $C_{\rm {rod}}$ for the triangular tubes ($C_{\rm {rod}}r_{\rm {rod}}\simeq 3.4$), 
one of the edges is not connected by the periodic boundary
and the rods assemble into a discoidal bud [see Fig.~\ref{fig:snap_n120}(b)].

The resultant tube shapes have hysteresis.
With decreasing $C_{\rm {rod}}$, the buckled discoidal tube can remain at $C_{\rm {rod}}r_{\rm {rod}} \gtrsim 2.85$.
Thus, the first-order transition between the triangular and discoidal tubes occurs
and the transition point is approximately $C_{\rm {rod}}r_{\rm {rod}} \simeq 3$.
Since the replica exchange method cannot be applied to discrete transitions with a large free-energy gap,
the exact transition point cannot be determined in the present simulation.
The triangular and discoidal tubes exhibit large amplitudes of $|r_{q\theta 3}|^2$ and  $|r_{q\theta 2}|^2$, respectively [see Figs.~\ref{fig:r_n120}(a), (b)]. 
Large amplitudes of $|r_{qz}|^2$ indicate buckling  [see Fig.~\ref{fig:r_n120}(c)].

\section{Simple geometric model of polygonal tubes}
\label{sec:geo}

To understand the dependence of the axial force $f_z$ and the formation of polygonal tubes,
we analyze the bending energy $F_{\rm {cv}}$ of the membrane tubes in simplified geometries.
The bending rigidity of the rods parallel and perpendicular to the rod axis
are denoted as $\kappa_1$ and $\kappa_2$, respectively.
The bending rigidity of the membrane in the absence of the rods is $\kappa_0=15k_{\rm B}T$.
When the rods are uniformly distributed in a cylindrical membrane tube
and completely aligned in the azimuthal direction,
the bending energy $F_{\rm {cv}}$ is expressed as
\begin{equation}
F_{\rm {cv}} = \frac{\kappa_1}{2}\Big(\frac{1}{R_{\rm {cyl}}}-C_{\rm {rod}}\Big)^2 \phi_{\rm {rod}}A
+  \frac{\kappa_0}{2}\Big(\frac{1}{R_{\rm {cyl}}}\Big)^2 (1- \phi_{\rm {rod}})A ,
\end{equation}
where $A=2\pi R_{\rm {cyl}}L_z$ is the membrane area.
The axial force $f_z= \partial F_{\rm {cv}}/\partial L_z|_A$ is given by
\begin{equation}
f_z = \frac{2\pi\kappa_0}{R_{\rm cyl}} +  2\pi\phi_{\rm {rod}} \Big( \frac{\kappa_1-\kappa_0}{R_{\rm cyl}} - \kappa_1 C_{\rm {rod}}\Big) . \label{eq:fz2}
\end{equation}
The first term is the force ($f_0$) in the absence of the rods.
The second term is the force generated by the rods,
which is proportional to $\phi_{\rm {rod}}$ and linear with respect to $C_{\rm {rod}}$.
These dependences agree with the simulation results (see Fig.~\ref{fig:vir_rho}).
Note that one should not take preaverages of the bending rigidity and spontaneous curvature in $F_{\rm {cv}}$
($\kappa= \kappa_0(1-\phi_{\rm {rod}})+\kappa_1\phi_{\rm {rod}}$ and $C_0= C_{\rm {rod}}\phi_{\rm {rod}}$).
If they are preaveraged,
the effects of the rods are overestimated and $f_z$ is not a linear function of $\phi_{\rm {rod}}$.
The bending rigidity $\kappa_1$ is estimated as $\kappa_1=40k_{\rm B}T =2.67\kappa_0$ from the fit of 
the slope of $f_z$ (see the dashed line in Fig.~\ref{fig:vir_rho}).
This value is reasonable for the parameter set of our simulation.
If hexagonal packing is assumed around an isolated protein rod, 
the protein particle has four neighboring membrane particles and two protein particles.
Although the coefficient of the bending potential along the rod is four times larger than that between membrane particles,
the effective bending rigidity is slightly reduced by the membrane--rod interactions parallel to the rod.
The absolute value of $f_z$ obtained by Eq.~(\ref{eq:fz2}) is smaller than the simulation results.
This is likely caused by the assumption of the complete azimuthal alignment of the rods 
and the negligence of the weakly attractive interactions between the rods.

\begin{figure}
\includegraphics[width=7.5cm]{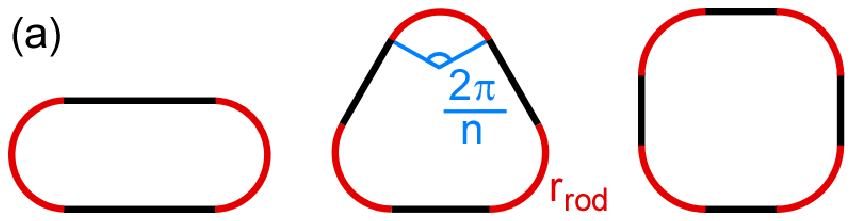}
\includegraphics{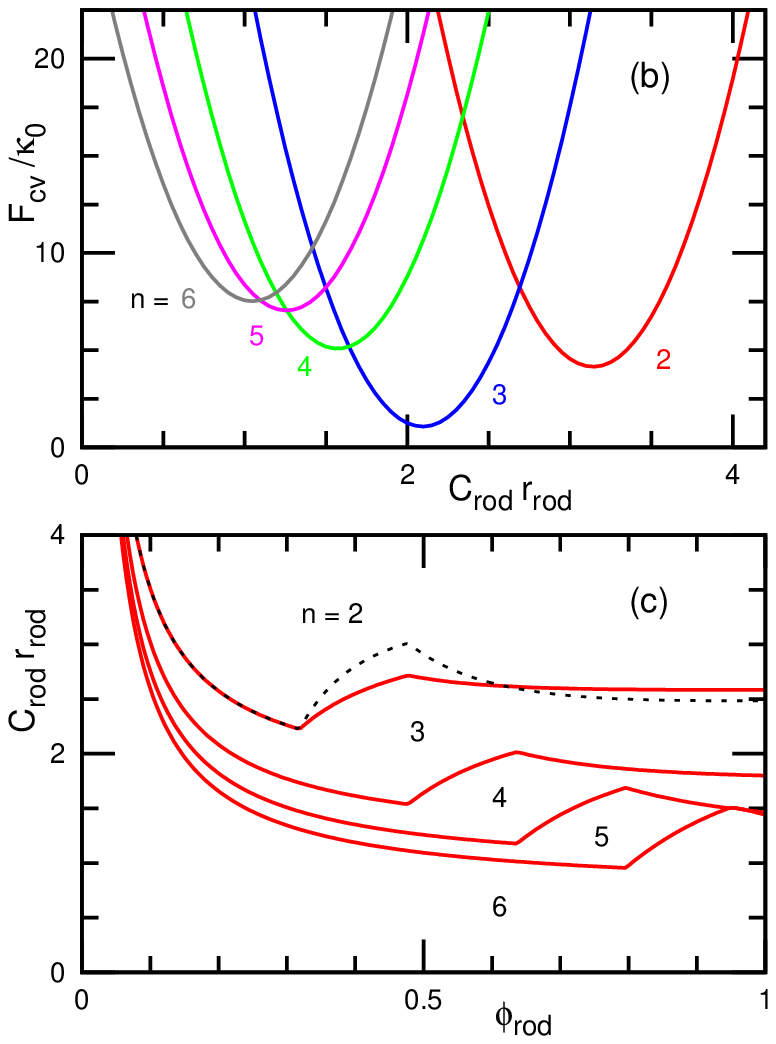}
\caption{ 
Membrane tube structures obtained by a simple geometric model.
(a) Schematic representation of slices of polygonal tubes with the number of edges $n=2$, $3$, and $4$.
(b) Curvature free energy $F_{\rm {cv}}$ of polygonal tubes 
at $R_{\rm {cyl}}/r_{\rm {rod}}=1$, $\phi_{\rm {rod}}=0.5$, and $L_z/r_{\rm {rod}}=4.8$.
(c) Phase diagram at $R_{\rm {cyl}}/r_{\rm {rod}}=1$.
The solid lines represent the phase boundary at  $L_z/r_{\rm {rod}}=4.8$.
The dashed line represents the phase boundary between the discoidal tube ($n=2$) 
and triangular tube ($n=3$) at  $L_z/r_{\rm {rod}}=2.4$. 
}
\label{fig:potube1}
\end{figure}

\begin{figure}
\includegraphics{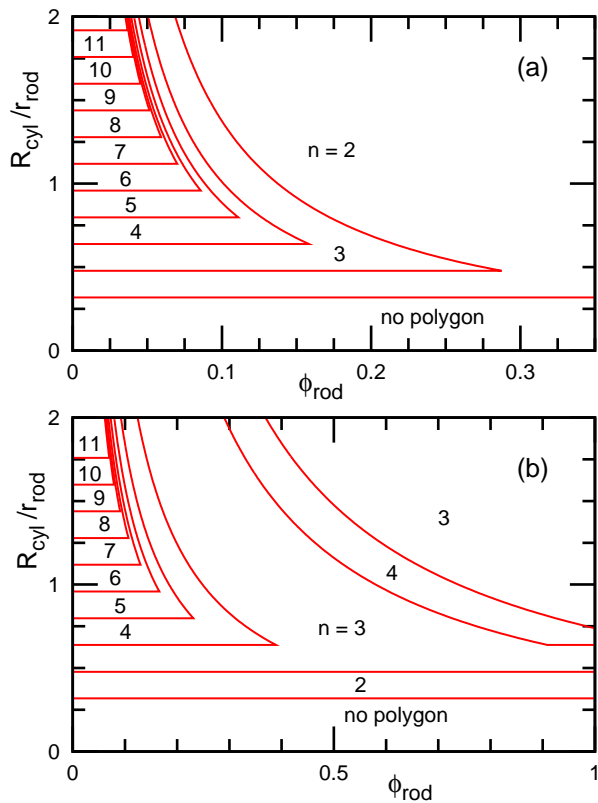}
\caption{ 
Phase diagrams of polygonal tubes obtained by a simple geometric model
for (a) $C_{\rm {rod}} r_{\rm {rod}}=3$ and (b) $C_{\rm {rod}} r_{\rm {rod}}=1.9$
at  $L_z/r_{\rm {rod}}=4.8$.
}
\label{fig:potube2}
\end{figure}

Next, we consider simplified geometries of polygonal tubes as shown in Fig.~\ref{fig:potube1}(a).
The $n$-gonal tube consists of $n$ flat membranes [black lines in Fig.~\ref{fig:potube1}(a)]
and $n$ round edges with a length $r_{\rm {rod}}$ and a curvature $2\pi/nr_{\rm {rod}}$ [red curves in Fig.~\ref{fig:potube1}(a)].
All of the protein rods are assumed to be aligned in the azimuthal direction and uniformly distributed in the edge regions.
The $n$-gonal tube can exist as a straight tube at $A\phi_{\rm {rod}}\leq A_{\rm {ed}}\leq A$,
where the edge area $A_{\rm {ed}}=n r_{\rm {rod}} L_z$ and the membrane area $A=2\pi R_{\rm {cyl}}L_z$.
The edges are buckled at $A\phi_{\rm {rod}}> A_{\rm {ed}}$.
The membrane does not have sufficient area to form an $n$-gonal tube
at $A_{\rm {ed}}> A$ ($R_{\rm {cyl}}/r_{\rm {rod}}<n/2\pi$).
The bending energy of the straight polygonal tube is given by
\begin{eqnarray} \label{eq:s1}
F_{\rm {cv1}} &=& \frac{\kappa_1}{2}\Big(\frac{2\pi}{n r_{\rm {rod}}}-C_{\rm {rod}}\Big)^2 A\phi_{\rm {rod}} \\ \nonumber 
& &+  \frac{\kappa_0}{2}\Big(\frac{2\pi}{n r_{\rm {rod}}}\Big)^2 (A_{\rm {ed}} - A\phi_{\rm {rod}}) .
\end{eqnarray}

The buckled polygonal tube also has non-zero curvature  along the longitudinal direction.
Here, we assume that the curvatures of the edge region are constant and 
 the curvature energy in the other principal direction (close to the azimuth) is approximated as Eq.~(\ref{eq:s1}) with $A_{\rm {ed}}=A\phi_{\rm {rod}}=nr_{\rm {rod}}L_{\rm {ed}}$,
where $L_{\rm {ed}}$ is the edge length and $L_{\rm {ed}}>L_z$.
The curvature energy along the longitudinal direction is given by
\begin{eqnarray}
F_{\rm {cv2}} &=& \frac{32\kappa_2 nr_{\rm {rod}}}{L_{\rm {ed}}}\Big\{(k^2-1)K(k)^2+E(k)K(k)\Big\}   \\
&=& \frac{32\kappa_2 nr_{\rm {rod}}}{L_{\rm {ed}}}\Big\{ \frac{\pi^2}{8}k^2 + \frac{3\pi^2}{64}k^4 + O(k^6) \Big\} \\ 
&\simeq& \frac{4\pi^2\kappa_2 nr_{\rm {rod}}}{L_{\rm {ed}}}\Big\{ 1-\frac{L_z}{L_{\rm {ed}}} + \frac{1}{4}(1-\frac{L_z}{L_{\rm {ed}}})^2 \Big\} , \label{eq:b3}
\end{eqnarray}
where $K(k)$ and $E(k)$ are the complete elliptic integral of the first kind and the second kind, respectively.\cite{nogu11a,comment_cor}
All of the edges are buckled with the same amplitude.
The modulus $k$ is determined from the length ratio,
\begin{eqnarray} \label{eq:Lx}
\frac{L_z}{L_{\rm {ed}}} &=& \frac{2 E(k)}{K(k)} - 1 \\ \nonumber
 &=&  1 - k^2 - k^4/8 + o(k^6).
\end{eqnarray}
For small buckling amplitudes, 
$F_{\rm {cv2}}$ can be expanded for $1-L_z/L_{\rm {ed}}\ll 1$
as in Eq.~(\ref{eq:b3}).

Figure~\ref{fig:potube1}(b) shows an example of $F_{\rm {cv}}$
obtained by Eqs.~(\ref{eq:s1}) and (\ref{eq:b3}) with $\kappa_1=2.67\kappa_0$ and $\kappa_2=\kappa_0$. 
At larger values of $C_{\rm {rod}}$, the polygonal tube with a smaller $n$ has the lowest energy.
The phase diagrams for $\phi_{\rm {rod}}$ {\it versus} $C_{\rm {rod}}$ at the simulation condition
are shown in Fig.~\ref{fig:potube1}(c).
The phase boundary is slightly underestimated 
($C_{\rm {rod}}r_{\rm {rod}}= 2.6$ in the simple model
and $C_{\rm {rod}}r_{\rm {rod}}\simeq 3.2$ in the simulation).
This deviation is caused by neglecting the translational and orientational entropies of the rods.

The polygonal edges are buckled at $\phi_{\rm {rod}}> n r_{\rm {rod}}/2\pi R_{\rm {cyl}}$.
At the buckling transition point, the phase boundary exhibits a kink. 
The buckling of a discoidal tube with $n=2$ starts at $\phi_{\rm {rod}}=0.32$.
The phase boundary between the $n=2$ and $n=3$ phases increases as the buckling amplitude increases with  increasing $\phi_{\rm {rod}}$
[see Fig.~\ref{fig:potube1}(c)].
As the triangular tube with $n=3$ is buckled at $\phi_{\rm {rod}}>0.48$, the boundary decreases again.
With a decrease in the tube length $L_z$, the buckling amplitude increases so that the phase boundary exhibits a greater increase,
whereas the boundary between straight polygonal tubes has no $L_z$ dependence
[compare the dashed and solid lines in Fig.~\ref{fig:potube1}(c)].
Reentrant phase transitions occur when the boundary of buckled and straight polygonal tubes is crossed with increasing $\phi_{\rm {rod}}$.

In experiments, protein density and tube radius are controllable parameters.
Figure~\ref{fig:potube2} shows the phase diagram for $\phi_{\rm {rod}}$ {\it versus} $R_{\rm {rod}}$ at fixed values of $C_{\rm {rod}}$.
As the tube radius increases, tubes with large values of $n$ can exist.
At large curvatures of $C_{\rm {rod}}$, the transitions occur between the straight tubes at small rod densities.
In contrast, transitions between buckled polygonal tubes can be more complicated at $1 \lesssim C_{\rm {rod}} \lesssim 2.5$.
At $C_{\rm {rod}}r_{\rm {rod}}=1.9$, reentrant phase transitions occur between $n=3$ and $4$ [see Fig.~\ref{fig:potube2}(b)].
With increasing $\phi_{\rm {rod}}$, a buckled triangular tube transforms into a straight square tube.
Subsequently, the square tube is buckled and transforms back to a buckled triangular tube.

\begin{figure}
\includegraphics{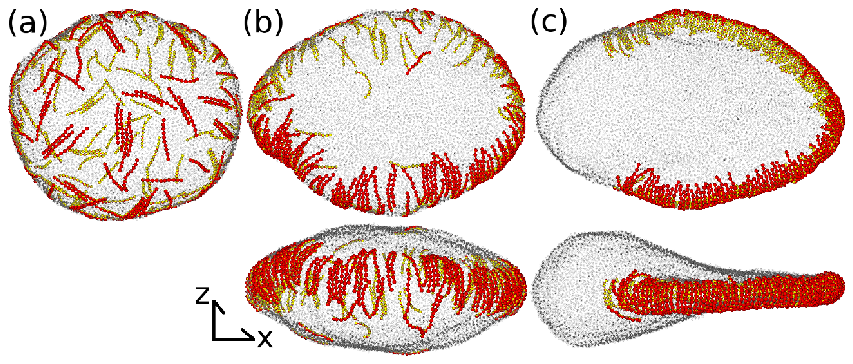}
\includegraphics{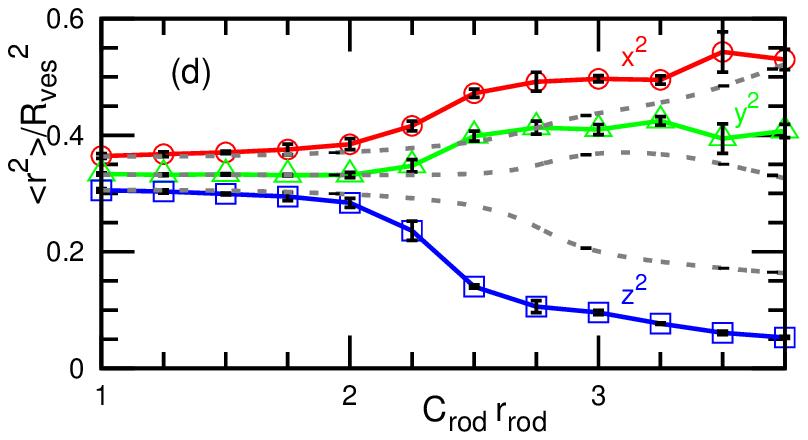}
\caption{
Vesicle shapes at $\phi_{\rm {rod}}=0.167$ for $N=2400$ and $9600$.
The snapshots are shown at (a) $C_{\rm {rod}}r_{\rm {rod}}=1.5$, (b) $2.5$, and (c) $3.5$ 
for $N=9600$.
Top and bottom snapshots are displayed in bird's-eye and front views, respectively.
(d) Three mean eigenvalues of the gyration tensor of the vesicle.
The dashed and solid lines represent data for $N=2400$ and $9600$, respectively.
}
\label{fig:n160}
\end{figure}

\begin{figure}
\includegraphics{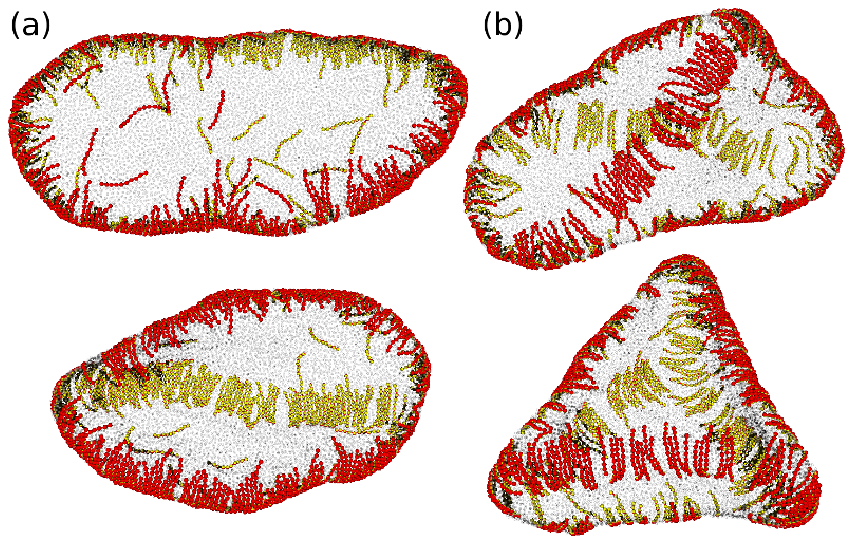}
\includegraphics{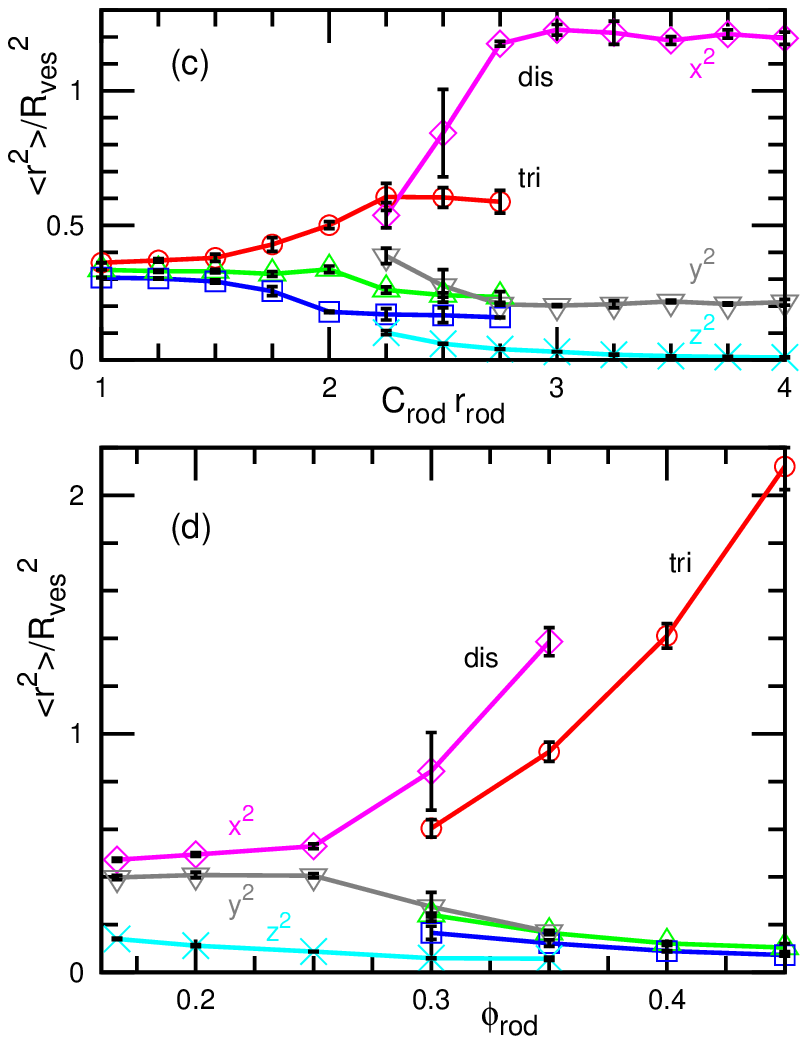}
\caption{ 
Vesicle shapes for various $\phi_{\rm {rod}}$ values at $N=9600$.
(a) Snapshots of vesicles of an elliptical disk and a triangular hosohedron at $C_{\rm {rod}}r_{\rm {rod}}=2.5$ and $\phi_{\rm {rod}}=0.3$.
(b) Snapshots of vesicles of a tetrahedron and a triangular prism  at $C_{\rm {rod}}r_{\rm {rod}}=2.5$ and $\phi_{\rm {rod}}=0.4$.
(c), (d) Three mean eigenvalues of the gyration tensor of the vesicle.
The labels 'dis' and 'tri' denote the vesicles with disk and triangular hosohedral shapes, respectively.
(c) Dependence on $C_{\rm {rod}}$ at $\phi_{\rm {rod}}=0.3$.
(d) Dependence on $\phi_{\rm {rod}}$ at $C_{\rm {rod}}r_{\rm {rod}}=2.5$.
}
\label{fig:n288}
\end{figure}

\section{Vesicles}
\label{sec:ves}

Self-assembly of the protein rods also occurs in a vesicle.
At small values of $C_{\rm {rod}}$, the rods are randomly distributed on the vesicle [see Fig.~\ref{fig:n160}(a)].
With an increase in $C_{\rm {rod}}$, the rods assemble into an equator of the vesicle (parallel to the rod axis)
and the vesicle deforms into an oblate shape [see Fig.~\ref{fig:n160}(b)].
With a further increase in $C_{\rm {rod}}$,
the rods also assemble along the equator (perpendicular to the rod axis).
Thus, the rods assemble in both directions,
and a discoidal bump is formed [see Fig.~\ref{fig:n160}(c)].
In our previous paper~\cite{nogu14}, we reported
the rod assembly at $\phi_{\rm {rod}}=0.167$ and $N=2400$.
Figure~\ref{fig:n160} shows that this assembly also occurs with large vesicles at $N=9600$.
These shape deformations are quantified by the changes of three eigenvalues of the gyration tensor
[see Fig.~\ref{fig:n160}(d)].
The gyration tensor is expressed as
$a_{\alpha\beta}= (1/N)\sum_j (\alpha_{j}-\alpha_{\rm G})
(\beta_{j}-\beta_{\rm G})$,
where $\alpha, \beta \in x,y,z$ and
$\alpha_{\rm G}$ is the center of mass.
When the vesicle is approximated as an ellipsoid,
these  eigenvalues are squares of the three principal lengths of the ellipsoid.
For the deformation into the oblate shape,
the smallest length $z$ decreases and the other lengths $x$ and $y$ increase.
For the bump formation, the largest length $x$ increases.

At high rod density $\phi_{\rm {rod}}$,
it is found that polygonal vesicles are formed in a manner similar to that of the polygonal tubes
(see Fig.~\ref{fig:n288}).
As $\phi_{\rm {rod}}$ increases,
the vesicle equator is completed filled by the rods
so that phase separation along the equator does not occur.
With a further increase of $\phi_{\rm {rod}}$,
the vesicle deforms into an elliptic disk instead of buckling
[see the top snapshot in Fig.~\ref{fig:n288}(a) and the increase of $x^2$ in Fig.~\ref{fig:n288}(d) 
from $\phi_{\rm {rod}}=0.25$ to $0.35$].
With an even further increase,
the vesicle transforms into a triangular hosohedron [see the bottom snapshot in Fig.~\ref{fig:n288}(a)].
The elliptic disk and triangular hosohedron coexist for finite ranges of $\phi_{\rm {rod}}$ and  $C_{\rm {rod}}$
[see Figs.~\ref{fig:n288}(d) and (c)].
These are the same types of behavior between discoidal and triangular tubes.
For $\phi_{\rm {rod}}=0.4$, the tetrahedron and triangular prism are obtained as metastable states.
The energy analysis in Sec.~\ref{sec:geo} suggests that 
these shapes and polyhedra with more edges can be stabilized at larger vesicles.

For polyhedra with genus $g=0$,
the number of faces $n_{\rm f}$, vertices $n_{\rm v}$, and edges $n_{\rm e}$
have the relation $n_{\rm f}+ n_{\rm v}-n_{\rm e}=2$ (Euler's formula).
In the polyhedra induced by the rods,
three edges are connected at any vertex.
This is different from elastic capsules \cite{lidm03,vlie06,bowi13} that prefer an icosahedron, where five edges are connected at the vertex.
Since each edge contacts two vertices,
the relations $n_{\rm e}=3n_{\rm v}/2$ and $n_{\rm e}= 3(n_{\rm f}-2)$ are obtained.
Fluid vesicles with linear defects~\cite{nogu03} also have this type of polyhedra.

\section{Summary}
\label{sec:sum}

We have revealed that the formation of polygonal tubes and polyhedral vesicles can be induced by 
banana-shaped protein rods.
At low rod density, rod self-assembly occurs in two steps.
As the rod curvature increases, the rods assemble in the direction parallel to the rod axis
and form discoidal tubes and vesicles.
Subsequently, the rods assemble along the disk edges.
As rod density increases,
the phase behavior remarkably changes.
With an increase in the rod curvature, 
the membrane tube transforms into a triangular tube 
and subsequently form a buckled discoidal tube.
The vesicle transforms into an elliptic disk and triangular hosohedral shapes.
The transitions between triangular and discoidal shapes are the first-order phase transitions.
These transitions can be understood by simple geometric analysis of the bending energy.

In our simulations, the number of protein on the membrane is fixed.
In typical experimental condition, the proteins are inserted from the bulk
so that the protein density on the membrane is determined by the balance of the chemical potentials.
Thus, in experiments, 
the protein density may discretely change accompanied by
 the shape transition between the polygonal tubes (or pohyhedral vesicles).
The polygonal tubes with more faces can be adsorbed by more proteins.
The drastic change of the the protein density may indicate a shape transition.

Polygonal tubes and polyhedral membrane shapes are often observed experimentally.
Regular arrays of triangular prismatic tubes are observed in the inner membranes of the mitochondria 
in astrocytes.\cite{sche08,blin65,fern83}
The formation mechanism of these triangular tubes is not known.
Our simulation study suggests that these prismatic tubes can be generated by the assembly of the BAR proteins.

In the presented simulation conditions, tubulation is not obtained.
Tubulation is likely prevented by the high bending energy at the tip of the membrane tubule,
resulting in linear rod aggregation at the edges of the polyhedra.
The addition of the spontaneous curvature of the membrane induces tubule formation.\cite{nogu14}
In living cells, the tubules and discoidal membranes are often connected.
Endoplasmic reticulum (ER) consists of flat discoidal membranes (rough ER) with branched tubular network (smooth ER).
It is an interesting problem to clarify the control mechanism of tubules and discoidal membranes
for further studies.
\\

\begin{acknowledgments}
The replica exchange simulations were
carried out on SGI Altix ICE 8400EX and ICE XA
at ISSP Supercomputer Center, University of Tokyo. 
This work is supported by KAKENHI (25400425) from
the Ministry of Education, Culture, Sports, Science, and Technology of Japan.

\end{acknowledgments}


\begin{thebibliography}{59}
\bibitem[{com()}]{comment_cor}
\bibinfo{note}{The sign of the second term in the bottom equation in Eq.~(7) of
  Ref.~\onlinecite{nogu11a}, should be replaced by '+'.}
\end{thebibliography}

\end{document}